\documentclass[12pt]{article}
\global\parskip 6pt

\normalsize

\begin{document}

\vspace*{1cm}
\thispagestyle{empty}
\centerline{\large\bf Measuring Consciousness}
\bigskip
\begin{center}
Siddhartha Sen\footnote{Email: ssen@tcd.ie}\\
{\em Centre for Research in Adaptive Nanostructures and Nanodevices,\\
Trinity College Dublin, Ireland.}\\
\end{center}
\vskip.5cm

\begin{abstract}
\subsubsection*{Abstract}
	In this paper a theoretical model for measuring consciousness based on experimental observations is proposed. Consciousness, the awareness of one's self and one's immediate environment, is defined, in this paper, in terms of the information about this awareness that is contained in or being processed by the brain. This particular information content in the brain is extracted from the correlation functions of brain waves observed in EEG measurements at $N$ points.  From these a time dependent probability function, $P_N(t)$ (Appendix B) is determined and then, Shannon's information theory formula,  $\sum-P_N( t) \ln P_N(t)$, is used to convert $P_N(t)$ to the information about awareness contained in the ensemble of brain waves (Appendix A).  The sum, in the formula, is over the $N$ points and the allowed range of brain wave potentials measured.  Consciousness is defined either directly by this formula $C(t)$, when it represents the information contained in brain waves at time $t$, or its time derivate $D(t)$, which itself represents the rate at which this information is being processed. These measures are not localized; they do not depend on a single, immutable, hardwired detail of the brain but they do reflect subjective experiences, encoded by any one of a number of possible hardwired circuits.  However, they are not clinical tools for measuring consciousness but rather represent ways to extract the information content of the brain from experimental measurements.  Justifications, based on observational evidence, are given for the formulas presented. 
\end{abstract}
\noindent
~~~~~~keywords: consciousness formula, correlation functions, brain waves, memory

\newpage

\subsubsection*{Introduction}
	Consciousness, our personal higher form of ``self-awareness'', is time dependent and has a limited focus of attention. For human beings it defines who we are.  The aim of this paper is to attempt to understand the nature of consciousness based on measurements made on the brain.  There is convincing evidence  linking consciousness to the brain see, for instance, Greenfield, 2001\cite{susan}.

	We relate consciousness to the information available in the neocortex of the brain at any moment of time, $t$, or to the rate at which this information is processed by the brain. These two related definitions give rise to two specific  formulas for measuring consciousness.  The relevant information content of the brain can be determined  from the measured brain wave EEG correlation functions by using  ideas from information theory (Appendix A) and a mathematical identity (Appendix B).  

	The first step required then is to justify the use of brain wave correlation functions to probe brain information. The second is to explain how the information content present in brain waves is extracted.  Thus we first discuss  the nature of the information contained in brain waves that have been established by experiment and
then then describe the procedure for extracting this brain information from brain waves.  Before we do this certain basic features of consciousness must be discussed.  
 
	Consciousness can be separated into two classes: primary consciousness and higher consciousness (see for instance Edleman\cite{edelman}).  Primary consciousness is our basic awareness of the world immediately around us and our knowledge of where our  body ends and the external world begins as well as our ability to distinguish between our friends and enemies.  This form of consciousness is necessary for all living creatures in order for them to survive and function.  In contrast higher consciousness involves  a thinking subject who is aware of his or her acts and also has an awareness  of mental constructions.  For this higher form of consciousness to operate requires the subject to have memory and the ability to  respond to external environmental changes.
 
	Thus in our measurement based approach we require that the variables selected have access to memory, that they respond  to external environmental signals and that they are time dependent.  The requirement for access to stored memories is crucial as it makes the approach sensitive to personal subjective experiences, thoughts and experiences.  The variables selected, the EEG correlation functions, are robust measurable time dependent  features of the biological brain that, as we show, have access to memory and are responsive to internal and external signals.  

	Many approaches for understanding consciousness have been proposed with very  different starting points and different objectives.  An incomplete list of references of some approaches are: non-linear  mathematics [Scott, 1995] \cite{scott}, quantum theory [Penrose et al, 1995, 2011]\cite{penrose}, quantum field theory [Vitiello  2001, Vitiello and Freeman, 2006, Stanford Encyclopedia of Philosophy, 2011]\cite{vitiello}, [Jibu et al., 1996] \cite{jibu}, [Chakraverty, 2014]\cite{chakraverty}, holography [Pibram, 2004]\cite{pribram}, evolutionary biology [Edelman, 1992]\cite{edelman}, neuoroscience [Crick, 1994,Fingelkurts, 2010, Fingelkurts, 2013]\cite{crick}, \cite{fing1}\cite{fing2} philosophy [Chalmers, 1995]\cite{chalmers} [Dennett, 1996; Putnam ,1988] \cite{dennett}\cite{putnam}, computer science [Churchland, 1986; Llyod, 2010] \cite{churchland}\cite{lloyd}, cognitive science  [Hobson,2000] \cite{hobson}, linguistics, psychology [James, 1977; Blackmore, 2002]\cite{james}  \cite{blackmore}, psychiatry [Hebb, 1980; Baars, 1988]  \cite{hebb} \cite{baars}, theoretical physics [Josephson, 1992] \cite{josephson} and religion [Nikhilananda, 2005] \cite{nikhilananda}.  Further references  can be found  in the works of the authors we have listed.  

	There are a growing number of works in which the ways of measuring consciousness has become the focus of attention.  A recent review  by Seth \textit{et al.} \cite{seth} examines some of  these  approaches and can be consulted for  an overview of the diverse methods used for this purpose.  The review includes approaches based on measuring EEG brain signals.  

	We briefly comment on two interesting approaches that use brain EEG data and  either endeavor to measure consciousness using statistical methods or suggest a way in which the structure of the data is related consciousness.  
 
	In the first approach due to Casali et al. \cite{casali}, the way in which brain waves are distributed and interact dynamically when probed by a transient magnetic field are  used to define operationally, a measure of consciousness.  The method has been clinically tested and found to be effective.  It can differentiate vegetative states from  minimally conscious states and can identify states in which a subject is conscious but cannot communicate due to motor defects.  However,  the approach does not suggest a theoretical interpretation of what constitutes consciousness.   
	
	The aim of the second approach, which has been studied by a number of authors, is to relate consciousness to the presence of hierarchical, temporal, architectural structures identified by brain waves.  (See, for instance, Fingelkurts et al., 2001 \cite{fing3}).  These transient structures are  shown to map onto structures and events within the outside world. It is suggested that by capturing these hierarchical, temporal, architectures a conscious robot could be built.  The idea of consciousness proposed in these works is time dependent and global but no overall, conceptual understanding of consciousness is proposed and neither is there any attempt to link consciousness to memory.  
 
	In a different direction Tononi [Tononi, 2008] \cite{tononi} has proposed a formula for measuring consciousness, based on an analysis of the structure and linkages between different operational subunits of the brain.  In this work consciousness is related to the structure of the biological brain by an explicit formula.  The mathematical formula is constructed using the idea of integrated information, which is a rule for determining the information present in the entire brain that is more than the information present in specially chosen subunits of the brain.  Consciousness, $C$, is defined as the difference between the information content of the entire brain, $E$, and that contained in the sum of a certain, prescribed set of brain subunits, $ \sum S_I$, that can be regarded to be independent unconnected entities.  Thus, $C$ measures the degree of entanglement present among brain subunits, which makes $E$ different from $\sum S_i$.  The quantity of integrated information present in a system is taken to be a measure of its state of consciousness, while the quality of a conscious experience depends on a "shape'' in the space of qualia.  The process of discarding informations introduced by Tononi [Tononi, 2008]\cite{tononi}  can be represented symbolically as the difference, $C \rightarrow E-\sum S_i$.  It  would be zero if the information content of the subunits, regarded as independent  entities, determined the information content of the entire brain.  

	The measure for $C$  introduced in this way is broad enough to be extended to discuss  the consciousness of  all entities whether they are living or non-living as all that is involved is the notions of units, subunits, entanglement and probability.  Hence consciousness is a universal property of the universe and levels of consciousness can be calculated.  Evaluating the consciousness value  for a human being at a given moment is, however, not easy. It requires  assigning probabilities and calculating the extent that these brain subunits deviate from a state of complete independence, based on a  knowledge of how they are linked together. Nevertheless, conceptually the framework of Tononi et al. represents a significant step forward because it introduces the idea of a complex system with subunits, all which are massively linked, as a model for understanding consciousness.  It also puts forward for the first time an operationally defined formula for consciousness.  

	However there is also a theoretical problem that Tononi's approach faces.  An insight of condensed matter theory is that knowledge of the structure of a complex system does not reveal its  function.\cite{vitiello}.  Consider a piece of matter.  If we are given all of the atoms and the way they interact with one another within the system we would not be able to predict that the system behaves as if it contains phonons.  Phonons, essential for understanding crystalline system, are not present as part of initial structure but, rather, only emerge once formation is complete. This observation means that identifying material brain structures with consciousness may not be possible. 
	
	In this paper a very different approach is used to derive two formulae for consciousness.  Unlike Tononi\cite{tononi} our approach does not depend on the hard wiring details of the brain; neither does it require identifying specialized subunits of the brain.  Instead, the approach proposes that consciousness for human beings, is always a time dependent feature that depends on memory and responds to external signals.  It is proposed that consciousness can be defined as either the information available in the brain at time $t$ or as the rate of processing this information.
	The key point of this paper is to describe and justify a procedure for extracting the information present in the  brain from EEG correlation measurements.The details of the procedure are given in the section where the model proposed is described.  Our starting point is the conjecture that the brain waves contain brain information. Brain waves observed in the EEG [Hobson,2000] \cite{hobson} represent the sum of all of the electrical pulses between masses of neurons communicating with each other within the brain (see  Freeman\cite{free2}, Tognoli \cite{tog}, Fingelkurts\cite{fing5}, Fingelkurts\cite{fing6}). Consequently, the dynamic state of the brain that can be measured at any point in time is represented by the total time-dependent information contained in the ensemble of brain waves at any given moment. This information includes that arising from emotions, facts, images, sounds, taste, memories and thoughts as well as information from the central, autonomic nervous system about the state of all our organs. This wide range of available information is contained in brain waves as the elegant experimental results of Colgin et al\cite{colgin} show that there is an ongoing dialog between brain waves and stored memory. 
			
	Our memory has a crucial property: it seems to be invariant over time, while our cellular proteins constantly change.  Thus,  memory provides the continuity of our self that we accept as a feature of our existence.  Our memory allows us to respond to changing situations, recognize friends, recall past events and engage in abstract activities, such as music, art, mathematics and science and define who we are.  Thus, any theory of consciousness should include memory as an essential feature.  This linkage is absent in the approach of Tononi\cite{tononi}. The  experimental study  of memory is currently undergoing a revolution see Tonegawa et al.\cite{tonegawa}, Queenan et al.\cite{queenan}, for a review of earlier work see Kandel et al\cite{kandel}. 
	
		A  basic observational  feature of the brain is that it operates in three well defined states.  These states are the waking state, the dreaming state and the state of deep sleep [Hobson, 2000]\cite{hobson}.  It has been found that associated with each of these states, there are specific, detectable, electrical brain waves ranging in frequency from less than 3 Hertz for deep sleep (delta waves), which are widespread, 4 to 8 Hertz for the dreaming state (theta waves) with a regional distribution that can involve many areas of the brain, to 8 to 12 Hertz for the normal awake state (alpha Waves), which usually involve the entire lobe with a strong concentration in the occipital region when the eyes are closed.  Frequencies above 13 Hertz to 40 Hertz represent a super alert state (beta and gamma waves) and these waves can be very localized.  All of the brain waves referred to above are present at all times but, depending on our state of wakefulness, one or other specific forms of these waves can be dominant.  For example, when a person opens his or her eyes, the alpha waves diminish and the beta waves increase.  However, in both of these states gamma waves are also present (see Baars\cite{baars}).  Recent research work has also established that whenever a person begins a new activity, a unique brain wave pulse with directionality is generated.  If the same activity is then repeated, the pulse observed is different from that observed during the first instance of the activity [Alexander, 2013] \cite{alexander}.  If brain waves do represent the state of the brain, this observation would make sense because the person themselves has changed between the two sessions of activity. 

	There are no brain waves constantly present in the cerebellum.  The ones that do at times appear there are in response to inputs and are in the theta-range, the gamma-range or are in a very fast oscillation range (over 80 Hertz).  Oscillatory behavior requires excitatory neurons.  In the cerebellum there are a few such excitory neurons in the form of granule cells.  Recent work [Knopfel, 2008]\cite{riken} has shown that the gamma and  the very fast oscillations are generated in the cerebellum when external agents are introduced but these oscillations do not come from the excitatory granule cells themselves.	Instead, they come from the inhibitory neurons that are present.  In addition the observed frequencies seem to  correspond to ones observed in the cerebral cortex under similar conditions of excitation.  

The process of reconstruction of the external world and the constant monitoring of the state of our internal organs  are both essential for us to live successfully in the world.  These brain activities are reflected in the brain waves observed. (Fingelkurt et al\cite{fing4}.  Brain waves change  when we open our eyes or hear sounds or when we faint.  
 \subsubsection*{What molds our consciousness?}
	We start life with the history of our ancestors encoded in our DNA.  This genetic information contains all of the mutations due to environmental mutagenesis and selection that our ancestors have experienced. At the penultimate end of this descent we have our immediate ancestors, namely our parents and grandparents.  This DNA history contains old data from the human species and its wanderings \cite{edelman, crick}.  Thus, at birth we arrive with certain predetermined physical characteristics and tendencies.  From then on the environment in which we grow begins playing a significant role.  By the environment we mean childhood upbringing, schooling, interaction with friends, books, films, music, social media, family, sports and so on.  There is also the well known effect of group and community behavior.  Group behavior seems to be generated by sound, non-verbal clues and the presence of some catalysts, all of which results in members of the group reacting in an atypical fashion, while community behavior is decided  by the norms and values of the community in which we live.  This behavior sometimes transcends one's own community and in special situations one either rejoices or mourns together as a nation or as a human being.  

 All of these observed effects have an effect on a person's memory, sense of 
belonging,  value system and consciousness.  This rich set of environmental
 effects listed does not pretend to be complete but it underlines the fact that defining
 who we are is complex.  Sometimes in this rich tapestry the influence of one
person or one event can determine the course of our life and sets the goals that we seek \cite{kac}.

    There are other important influences in our life that do not come from the environment or from interactions with others but, rather, from more abstract sources.  For instance there are books and religion that tell us what are the aims of life and how it should be lived.  Then, there are legends and histories of our country that very often suggest certain codes of conduct, certain aspirations and certain ways of regarding those not belonging to our own country, clan or community.  They are, thus, of great importance for our perception of who we are \cite{radley}\cite{wasserman}\cite{schlitz}. 

	The changing nature of our consciousness with age is reflected in the distribution and nature of our brain waves \cite{craik}.  As our consciousness is molded our brain waves change. These remarks suggest that consciousness is not determined simply by
the hardwiring  of the brain but on the ever changing coherent dynamics within an assembly of brain cells. Our awareness is influenced by our past experiences and our
environment. It is a coming together of all our sense and emotional inputs including
the state of our health. Conscious states are thus composed of representations of the world at the present moment and embedded within a representation of what is present are the sensations, feelings and thoughts that represent the past \cite{doege}.

   Let us now describe our model.
  \subsubsection*{Model for Consciousness}
  We have proposed a measurement-based approach for the study of consciousness. A
  measure of a conscious experience is defined as the dynamic information contained in the brain at any given moment of time $t$ and the measure of consciousness as the rate at which this information is processed at time $t$. These measures require time dependent brain wave observations. We have chosen EEG brain wave correlation functions as an example of an appropriate set of variables to use and have justified this choice using experimental results. The ensemble of brain wave correlation functions have two essential features. They contain subjective memory information \cite{colgin}and they  reflect and respond to both the external world and our complex  inner world \cite{fing4}. 
 
 We also note that consciousness has different degrees.  The lowest degree of consciousness distinguishes between the boundaries of living animals and their environment and allows an animal to identify food and avoid predators. The higher degrees of consciousness lead to the complex notion of self awareness,
 necessary if a person is to have aspirations, to experience love, compassion, self sacrifice as well as a feeling of self worth.  We have suggested that this higher level of consciousness crucially depends on memory, that in turn depends on many things, including our environment.  Finally, we have suggested that besides the reasonable channels of memory formation, such as through reading, hearing, personal experiences and human contacts there could be further, presently unknown sources, coming from our genetic history.
    
   We show (Appendix B) how a time dependent probability function $P(t)$ can be constructed from the measurement of brain wave correlation functions.This function, $P(t)$, has the  property that the space average of brain potentials using it reproduces the measured brain wave correlation functions .  Consciousness, $C(t)$, as the information content of brain waves at time $t$, is then, according to information theory,  given by the formula $C(t)=\sum -P_N(t)\ln P_N(t)$.  This is the Shannon information formula (Appendix A) for brain waves and defines for us the dynamical brain. 
Alternatively, consciousness can be defined as the rate of processing this information $D(t)$ given by $D(t)=\frac{dC(t)}{dt}$.  Thus, consciousness in the model represents the information content of or the rate at which information is processed by the dynamical brain at time $t$.  It is relatively easy to justify choosing EEG data for understanding consciousness because there is overwhelming evidence establishing that brain waves reflect our interactions with the external world (see, for instance Freeman\cite{free1},Nunez et al\cite{nun},Fingelkurts\cite{fing4}).  Brain waves also reflect our conscious and even our subconscious thoughts as well as our personal, subjective experiences.  This realization follows from the fact that brain waves interact with memory and hence have access to our conscious and subconscious thoughts and our stored experiences [Colgin et al, 2009]\cite{colgin}.

  We should note that the measures of consciousness proposed are based
  on extracting the information contained in the brain from brain wave measurements.  They are not meant to be used as clinical tools as proposed by Casali\cite{casali}.   
  It should also be stressed that the specific measures proposed certainly do not capture the full richness of conscious experience. They are, however, the start of a project to study subjective consciousness using brain measurements.  

      	We now give details of  how to use measured brain waves to measure consciousness.  Consider the measured correlation functions, $G_{12}=G(V(x_1,t) V(x_2,t))$, for points $(x_1, x_2)$ of  the brain, at time, $t$, to represent the state of the dynamic brain at time, $t$, where, $V^{a}(x_1,t), V^{b}(x_2,t)$ represent measured electric field values at two points for waves, $a, b$, that can represent any one of the observed brain waves, {\em i.e} the alpha, beta, gamma, theta or delta waves.  We will suppress labels, $a, b$,  present in $V(x,t)$, from now on and use the notation, $V_i$ for $V(x_i,t)$ and $G_{ij}$ for $G( V_i V_j)$.  Each set, $V_1 V_2..V_N$, represents a message of length, $N$.

Using standard tools of mathematics we show, in Appendix B, that 
from the measured brain wave  correlation functions, $G_{ij}=G(V(x_i,t)V(x_j, t))$,
a time  dependent probability function can be constructed. It is:
\begin{eqnarray*}
P_N(t)&=&\frac{1}{Z}e^{\sum_{m,n+2}^{N}[V_m(G^{-1})_{mn}V_n]}\\ Z&=&\frac{\sqrt{1}}{\sqrt{|det(G^{-1})_{mn})|}}
\end{eqnarray*}
where, $|det(G^{-1}_{mn})|$, is the determinant. This is the key step of our approach. Once $P(t)$ is constructed from data we can immediately 
 define the information content of brain waves at time $t$ by the Shannon information theory formula as,
 \begin{displaymath}
   C(t) = -\sum P_N(t)\ln P_N(t) 
\end{displaymath}   
Shannon's formula is discussed briefly in Appendix A. The sum is over the $N$ points
measured and the range of allowed values of the potentials $V_m$ is measured at time $t$. The range of these variables is $-20<V<100$ in units of millivolts. We identify $C(t)$ as a measure of a conscious experience and $D(t)=\frac{dC}{dt}$, as a measure of consciousness. It represents the rate of processing information.
 The formulas proposed are theoretically well founded and grounded in data: they are not an ad hoc  constructions. Both $C$ and $D$ include subjective experiences.

 The measure is chosen to have two properties: $C(0)=0$ and
  $(\frac{dC(t)}{dt})_{t=0}>0$ These are important  requirements. The absence of 
time dependence in the brain waves signals death and, hence, the absence of
 consciousness. For this reason we set $C(0)=0$. When time dependence 
 is present consciousness increases from zero hence $(\frac{dC(t)}{dt})_{t=0}>0$. Thus in our approach consciousness of a human being cannot be constant. It must change with time. 
 An immediate consequence of these conditions is the result for $t$ small:
 \begin{eqnarray*}
 C(t)&\approx& C(0)+t\frac{dC}{dt}\\
 &=& t\Omega
 \end{eqnarray*}
 Thus for small times $C(t) \approx t\Omega$ where $\Omega$ is a characteristic
 frequency associated with a brain wave. We have taken $C(0)=0$. In general 
 a response to an input signal corresponds to a change in $t$ from $t_0$ to
 $t_0+t$. Then we have $C(t_0+t)-C(t_0)=t\Omega$. Thus $C(t)$ can increase
 or decrease depending on whether $\Omega$ is larger or smaller than it was 
 at time $t_0$. 
 
Let us suppose we can introduce an effective probability function of the form,   $P(t) \approx e^{-F(t\omega)}, F(\omega)>0$. Such a structure is reasonable from  dimensional reasoning.  The formula gives $C(t) \approx \frac{dF(t\omega)}{dt}e^{- F(t\omega)}$ where we have used  the fact that the dominant contributions come from the region where $F(t\omega)<1$. This simple example is not meant to be realistic as we have used a single effective probability with just one frequency $\omega$.  Nevertheless, this example makes the link between frequency and $C(t)$ clear.   In the next section we give a quantitative argument linking different states of awareness and frequencies..

\subsubsection*{Concluding Remarks and Comments}
An approach to the study of consciousness, based on experimental measurements
of brain waves (neural oscillations) is proposed. In the approach a method for extracting the information present in brain waves is suggested. This involves
constructing a probability function $P(t)$ from the measured brain wave correlation functions with the property that the average value of brain wave potentials
calculated using it, namely, $<V(x,t)V(y,t)>=\int dV P(t,V) V(x,t)V(y,t)$, give the measured correlation function between the chosen points $x, y$ (Appendix B). 

  Information theory tells us that if the probability, $P$, for a message of length $N$ is known then the information contained  in all possible messages of length $N$ received is given by a formula due to Shannon \cite{shannon}.  Here we have taken a given set  of measured brain wave potentials $V_1V_2..V_N$ at time $t$ to represent a message of length $N$. All possible messages of this length then correspond to the allowed range of values of  the potentials $V_i, i=1,2,..N$ and on the range of values  of $N \leq N_m$, where $N_m$ represents the total number  of measured points.  The Shannon formula, $C(t)=\sum_V-P\ln P, i=1,2,...N$, can be used to determine the information contained in all brain wave messages if we have a theoretical method for assigning a probability to a given brain wave message $V_1,V_2,..V_N$ of length $N$. In Appendix B we explain how such a probability function can be constructed from measured EEG correlation functions. These correlation functions contain the effect of all the complex interactions present between brain waves as they are  based on measurements, not theory. They  are not expected to have any universal structure but to have forms that reflect the complex interactions between brain waves at the moment of time  when observations were made. Thus the probability function obtained is not a theoretical  construct but based on data.

A measure of consciousness $D(t)=\frac{dC}{dt}$ is then defined as the rate of change of the brain wave information at time $t$t. The definition it should be stressed is a specific measure of consciousness. We do not think a simple function of time can completely  capture the richness of experienced consciousness. Such a proposal is useful if clinical  or neuroscience evidence is provided to establish that brain waves have three important properties. 
 
   The first property is that brain waves provide an accesible means for exploring the dynamical state of the brain. This is supported by neuroscience  research \cite{koch}\cite{linas}. A quote from Northoff\cite{northoff} makes the point well.  ``In the same way as physical activity can be relationally determined, the brain can encode neural activity in a relationally determined way.''  

  EEG measurements are one way to measure the brain's neural activity. The method has its technical limitations as scalp measurements cannot probe subcortical neural  properties.  However there are a variety of  tools available to study neural oscillations
  and neural correlation functions such as fMRI, ERP, PET, MEG, etc \cite{huber}. 

The second property is that brain waves constantly interact with the memory centers of the brain. It is well known that brain waves play an important role during memory formation \cite{kandel, ryan}, however, to establish the presence of a constant interaction between brain waves and memory centers is important as it implies that brain waves contain in them  personal subjective experiences. Supported for such a feature comes from the observations of Colgin et al \cite{colgin} and others \cite{miller, amir}who found that brain waves carrying incoming information, access memory centers of the brain many times a second. This is done in order to interpret the nature of the in coming  information . The process of comparison with memory creates ``awareness'' in a person regarding the nature of the incoming signal. As memories are subjective we can conclude that brain waves carry subjective information.  It should be stressed that the brain wave patterns at any given moment of time, contain only a small fraction of the available information in the dynamic brain. The nature of this information is dependent on the areas of the brain that are probed, the time and the environment.

The third property is that there is strong clinical evidence linking memory and consciousness. There are numerous works on this topic\cite{baars2, endel2,ebbinghaus, gardiner, doege, shevlin}. Tulving\cite{endel2} provides empirical observational support for links between different memory systems(procedural, semantic, and episodic) and corresponding varieties of  consciousness  ( anoetic, noetic, and autonoetic) they are related to. In particular that episodic memory has autonoetic consciousness as its necessary correlate. Procedural memory is included in Tulving's list, but this form of memory does not contribute to consciousness in the sense of ``awareness''. Thus the consciousness associated with this form of memory Tulving calls, anoetic  (not knowing) consciousness.  Implicit memories that do not contribute to awareness  include the self regulatory centers of a living system that regulate breathing, heart beats and body temperature \cite{ledoux}. In the absence of these memories there  would be no consciousness in a human being. Memory, as discussed by Tulving and others cited, is thus  necessary for consciousness. But consciousness and memory are not the same.  Consciousness is a  moment-to-moment feature of living systems, while memory  provides the continuity of a person's self, contributes to the content of conscious awareness, but it also provides the information necessary for a person to survive in the world.

We can conclude that the measure of consciousness proposed is useful and that it contains subjective information. The  hard problem of consciousness for the model is thus to decode brain wave information, that is to understand, for instance, the way brain waves encode memory information and information regarding our vital organs.

 The work of Fingelkurts et al.\cite{fing4} is an important step in the direction of decoding brain wave information. They show how structural brain wave patterns can be mapped to events of the external world, but the imprint of memory present is not studied.
 
   In our  brain wave picture the quality and meaning of a conscious experience is captured not just by $C(t)$  but by the unfolding in time of  a three dimensional surface representing the distribution  of $V_i, i=1,2,..N$ values as a result of an incoming signal  interacting with memory.  The unfolding time is expected
 to be a fraction of a second rather than milliseconds\cite{colgin}.

 Finally we briefly address a conceptual issue regarding the way consciousness and information are related in the model. The basic idea of the model is to construct a specific measure of consciousness in terms of the  information content of the brain, extracted from measurements.  We used measured brain wave information to
 to access the information content of the dynamic brain and used this procedure to construct our measure of consciousness. Such a hypothesis was made  because of the link, supported  by clinical  research, between brain waves and  memory and between memory and consciousness \cite{endel2}. Memories  contain subjective information hence the hypothesis that consciousness is related to the rate at which  the brain processes information is reasonable. The information relevant for human consciousness was associated with the brain, was time dependent, had the ability to constantly change in response to external input signals and was embedded in an interactively generated memory system \cite{doege,endel2}. Consciousness is a feature of a system.  However the model can determine the degree of consciousness   given any information function $C(t)$ by simply determining $D(t)=\frac{dC}{dt}$ but the value obtained will be interpreted as the degree of consciousness of a system that has  $C(t)$ as its information function (Appendix C). It should be clear  that a specific measure of an attribute is not the same as the attribute. Thus abstract information without further qualifications is not conscious \cite{pockett}. 
 
 Our approach follows the standard practice of science, namely a system is specified (the brain), the abstract idea of information contained in the system is determined  from observation, using mathematical ideas, and then a specific hypothesis is made
 linking the rate of processing brain wave information to human consciousness.
 This hypothesis can be tested. The test is to check that the degree of awareness of a person is related to the frequency of brain wave oscillations. We will discuss this result shortly.  For modeling a physical system a similar sequence of steps are followed. A system of interest is identified, a suitable set of relevant abstract ideas like that of an electromagnetic field or of energy or of entropy are identified and extracted from measurements with the help of theoretical ideas and then the abstract ideas are used to draw testable conclusions regarding the system. For example the notion of entropy listed, underpins statistical mechanics, used to study the thermal properties of systems.  The abstract concept of entropy can be formulated using  information theory \cite{harris}. For systems in thermal equilibrium 
 the information content is time independent and hence the system's consciousness is zero.

 The measure of consciousness proposed has testable consequences. We stated, for example, that  the measure predicts that there should be a positive correlation between brain wave frequency and the degree of consciousness.  This prediction follows from the formula proposed as we will show. However at the observational level it is known that a person in deep sleep, in a coma or with a brain injury have, predominantly, very slow delta waves, leading to a small value for $D(t)=\frac{dC}{dt}$, while states with higher brain wave frequencies, representing higher levels of awareness, will have larger values for $D$. This statement requires  clarification.  In REM sleep both high frequency gamma waves and low frequency theta waves are present.  However gamma waves are also present when alpha and or beta waves occur.  Thus, $D(t)$ for REM sleep is lower in value than in those states where alpha or beta waves are present. But the REM sleep value for $D(t)$ is higher than it is for a person undergoing an epileptic absence seizure  even though in this state there is high EEG activity.  This dicotonomy is observed because the brain waves during an epileptic absence seizure have high amplitudes but low frequencies (Seth et al.\cite{seth}).  We can take states with $D(t)$ values lower than or equal to those obtained for REM sleep to represent states of unconsciousness.
 
         Let us provide a quantitative version of these remarks. Suppose $C(t)=\sum _{i=1}^{5} c_i C_i(t)$ where the sum represents the contributions  to consciousness from an ensemble of brain waves  that have delta $ (i=1)$, theta (i=2), alpha (i=3), beta (i=4) and gamma (i=5)  brain waves in them, with probabilities $P_i, i=1,2...5$. In this expression the average over allowed potentials is
implimented directly by a single probability for each type of brain wave. We will call this replacement the mean field approximation. The numbers  $c_i, i=1,2..5$ are weights, constant numbers at a certain period of time, that satisfy two conditions: $c_i \geq 0, \sum_{i=1}^{5}c_i=1$. In any given moment of time all the waves are present but the value of the weights depend on the state of awareness. Thus in the state of deep sleep $c_1$ dominates while in the awake state $c_4$ dominates. In states $i=2,3,4$ there is a $c_5$ i.e  a gamma wave, component present. We next  note that $C_i(t)$ and $c_i$ are dimensionless numbers as they are related to  probability functions.  For $C_i(t)$ to be dimensionless requires it to be a dimensionless function of $t$, namely $\omega_it$. From this it follows that $D(t)=\frac{dC}{dt}$ has the structure $D(t)=\sum_{i=1}^{5}c_i\omega_i \frac{dC_i(x)}{dx}, x=\omega_i t$, where we use $\frac{C_i(\omega_i t)}{dt}=\omega_i\frac{dC_i(x)}{dx}, i=1,..5$  It is shown in Le Van Quyen et al\cite{quyen} that gamma waves are always present in the wake-REM sleep cycle i.e the factor $c_5$ is roughly the same for these states while it is absent for the epileptic absence seizures.  Thus the level of consciousness for
an epileptic absent seizure is lower than it is in REM sleep although there is considerable neural activity present. The sketch given makes a number of simplifying assumptions but it clearly links $D(t)$ to brain wave  frequencies.
    
 Another important consequence of our approach is that it explicitly 
notes that consciousness and life depend on the presence of both external and
 internal time dependent sources that are responsible for the observed
 time dependence of brain wave signals. This observation means that consciousness is not determined  simply by the hard-wiring of a person's brain  but, rather on many other factors.  
    
 What is the self in the model?  We have provided an operational definition, based on experimental results, that the self initiates our thoughts and actions in a way we do not know and that it provides us with our sense of being the same person over time.  This notion of continuity of the self comes from our memory since the individual copies of our basic cellular proteins are degraded and replaced over a period of a few weeks while our sense of self persists for all of our normal waking lives. There is also a second observational feature that we describe by the term, "self''.  This feature arises from the work of Libet\cite{libet},  which shows that an action we initiate seems to start before we become aware of having initiating it.  We describe this result and our feeling of the continuity of our personality as a defining feature of our "self''. This form of the term, "self", as we use it in this instance, is a convenient terminology. For a glimpse of the many dimensions of ``Self'' , the works
 of Eccles \cite{eccles}and Ramachandran \cite{ramachandran}maybe consulted.

	We end by saying that, although we have proposed a measurement based approach for consciousness that includes subjective data,  we still do not know what it is or where it is located.  We still do not know what seeing or hearing really means, nor do we know what friendship, love, anger, loyalty or beauty are.  These are all words that represent experiences and emotions.  Perhaps the best we can do in our attempt to understand consciousness is to try to capture important aspects of its abstract core by testable measurements.  The  remark regarding the absence of real understanding also holds for matter.  We do not know for instance what an electron "really'' is.  All that we can say is that it can be characterized by a set of properties that can be measured and that we can predict the way it behaves in different situations.  
 
\subsubsection*{Acknowledgement}
I would like to thank  Paul Voorheis for discussions, useful comments, and for extensive editorial help,  Mani Ramaswami and Mike Coey for many discussions, comments and encouragement  and  Jean-Pierre Gerbaulet for a careful reading of the paper, for comments and for drawing my attention to the work of R.Kurzweil on thinking.

\subsubsection*{Appendix A: Basic Formula of Information Theory}
Any device, system or process for generating messages is called an
information source.  Each source has an alphabet. 
A message containing information has regularities. It has statistical structure.
 A real source of information can be modeled, for example, as we do, by  
 time dependent correlation functions, which contain a messages formed from the alphabet given by the measured values of the electric field $V_i$  at large number of different points of the brain. 
These messages contain information that are of interest.  If a time dependent probability function  $P(t)$  can be assigned to such a message then  $ A=-\sum P_n \ln P_n$ is the average information per message.  For $P(V_1,...V_n)$ the sum is  over the location and range of values of all the $V_i$ present in $P${\em i.e} can take. 
 The maximum value  this sum $A$  can take is called the capacity, $K$ and
 is given by $K=M \ln N$, where $M$ is the range of values the variables $V_i$
can take and $N$ the total number of source  points
possible. For the brain we can take $N \approx 10^{14}$  then 
$K \approx M\times 50$ bits. For visual perception 
the maximum rate of information processing is found to be in the range 
of $40$ to $50$ bits per second [Van Essen, 1991]\cite{vanessen} while the information collected by the photoreceptic mosaic of the eye is $\approx 10^6$ \cite{jacobson}.  

 \subsubsection*{Appendix B: Assigning Probability to Correlation Functions}
 Consider correlation functions  of electrical fields $G_{i j}=G(V_i,V_j)$ at any two  brain points measured at times $t_i. t_j$ for measured potentials  $V_i=V(x_i,t_i), V_j=V(x_j,t_j), j=1,2...n$.  We suppose that measurements are made  for  a large number of point  $i_1, i_2,...i_N$, and  take the set of these measurements to represent a state of the dynamical brain at time $t$ when $t_1-t_2,..t_N=t$. We 
 also suppose $G_{ij}=G_{ji}$. The set of values  $V_1,V_2,..V_N$ at time $t$ is a message of length $N$ in the sense of information theory. The set of all messages then corresponds to all  allowed  values for this set for $1\leq N \leq N_{max}$,
 where $N_{max}$ is the maximum number of points where measurements have been made. We assume that the eigenvalues of the matrix $G_{ij}$ are real and positive. This reflects the absence of strictly localised brain waves. However the model
 can be adjusted to deal with such excitations if EEG data reveals their presence.
 
 There is a simple procedure  to extract a probability
 function from correlation functions. We start from the set of measured 
 $G_{ij}$ values and use the following  identity,
 \begin{eqnarray*}
 G_{ij}&=& \frac{1}{Z}\int [\frac{dV_1}{2\pi}\frac{ dV_2}{2\pi}...\frac{dV_N}{2\pi}] e^{-\frac{1}{2} \sum_{m,n=1}^{N}  [V_m( G^{-1})_{mn}V_n]}  V_i V_j  \\
 Z&=& \int [\frac{dV_1}{2\pi}\frac{ dV_2}{2\pi}...\frac{dV_N}{2\pi}] e^{-\frac{1}{2} \sum_{i,j=1}^{N}[V_i( G^{-1})_{i j}V_j]}
 \end{eqnarray*}
 where $\int [\frac{dV_1}{2\pi}...\frac{dV_n}{2\pi}]$ represent integrations over the measured set  of numbers $V_1...V_n$. These numbers all lie in a given range of values defined for the each  type of brain wave considered. However for the identity to hold the range of any $V_m$ is $-\infty<V_m<+\infty$ and $det(G_{ij}) \neq 0$. This replacement of finite ranges by infinite ranges does not introduce a large error. .  The time dependences present in $V(x,t)$ is due to  external and internal sources of information which are space and time dependent. 
 
 Let us consider the case of where only  a single pair of 
measurements $V_1=V(x_1,t_1), V_2=V(x_2,t_2)$ is used to describe the brain state. 
 The  probability $P$ of this brain state, in the model, is given by (setting $t_1=t_2=t$,
 \begin{eqnarray*}
 P_{2}(t)&=&  \frac{1}{Z}e^{-\sum \frac{1}{2}[ V(x_i,t)(G^{-1})_{ij}V(x_j,t)]}\\
 Z&=&\frac{1}{\sqrt{| det (G^{-1}( V_i,V_j)|}}
 \end{eqnarray*} 
 Here $det |G|$ is the determinant function of the measured correlation function $G$ and  $i, j$ can take the two value, $1,2$.  Then $G_{1 1}=<V_1 V_1>, G_{2 2}=<V_2 V_2>, G_{1 2}=G_{21}=<V_1 V_2>$,  where $<V_i V_j>$ is the two point correlator at time $t$. For the case when $N$ different point measurements  are made for times $t$, we have the $n$  point correlation function $G(i_1,i_2,...i_N) $ where $V_1=V(x_1,t), V_2=V(x_2,t)...,V_N=V(x_N,t)$ can be similarly expressed in terms of measurements. The corresponding probability function is
 \begin{eqnarray*}
 P_{N}(t)&=& \frac{1}{Z} e^{-\frac{1}{2}\sum_{m,n=1}^{N} [V_m (G^{-1})_{mn} V_n]}\\ 
 Z&=&\frac{1}{\sqrt{|det (G^{-1}_{mn})|}}
 \end{eqnarray*}
 In order for our analysis to apply the matrices $G$ must have non zero determinant.  
 \subsubsection*{Appendix C: Consciousness of Matter}
 Consider a collection of hydrogen atoms at temperature $T$. The states of
hydrogen atom are known from quantum theory.  Hydrogen atom states are  
described by a discrete set of oscillating frequencies $\omega_n$ and 
  probability wave functions.  They are  $ \omega_n=\frac{\omega_0}{n^2}$ 
where $n$ can take only discrete value{\em i.e} $n=1,2,3...\mbox{infinity}$ \cite{dirac}.
The probability $P_n$ of hydrogen
 atoms at temperature $T$ to be in the state with frequency  $\omega_n$  is,
 from statistical mechanics [Landau and Lifshitz, 1978] \cite{landau}, given  by 
 \begin{eqnarray*}
P_n&=&(2n+1)\frac{e ^{-\alpha_0\frac{\omega_0}{n^2}}}{Z}\\
Z&=&\sum (2n+1) e^{-\alpha_0 \frac{R}{n^2}}
\end{eqnarray*}
where $\omega_0 $ is known positive constants and $\alpha_0=\frac{\hbar}{kT}$,
$k$ is the Boltzmann constant. The sum is over all discrete values of $n$.
 Our formula for consciousness $C$ is
\begin{displaymath}
C=\sum_{n=1}^{\infty} - P_n \ln P_n
\end{displaymath}
But $P_n=0$ for hydrogen as $Z =\mbox{infinity} $. Thus $C=0$. However this
 calculation is flawed as all the discrete set of values used do not exist at temperature $T$. However as $C$ is time independent the system has zero consciousness $D=\frac{dc}{dt}$.

  For classical planetary orbits each orbit has probability $P=1$ and hence the system has $C=0$.  This result will be  true for all deterministic systems of classical physics as $P=1$ for any deterministic system. 

For a single hydrogen atom with discrete frequency $\omega_n$ the associated
space averaged probability $P_n=1$ for each value of $n$. This result follows from
 quantum mechanics as the discrete frequency states of hydrogen atom are all
 stationary state {\em i.e} the probabilities of these states are time independent. Hence $D=0$
 \end{document}